\documentclass[sigconf]{acmart}
\usepackage{pifont}
\usepackage{verbatim}
\usepackage{subfigure}
\usepackage{algorithm}
\usepackage{algpseudocode}
\usepackage{pifont}
\usepackage{graphicx}
\usepackage{subfigure}
\usepackage{multirow}

\usepackage{tabularx}
\usepackage{enumitem}
\usepackage{fancybox}
\usepackage{listings}
\usepackage{xcolor}
\usepackage[most]{tcolorbox}
\usepackage{framed}
\usepackage{changepage}
\usepackage{verbatim}
\usepackage{algorithmicx}
\usepackage{soul}

\definecolor{shadecolor}{rgb}{0.9, 0.9, 0.9}
\definecolor{codegreen}{rgb}{0,0.6,0}
\definecolor{codegray}{rgb}{0.5,0.5,0.5}
\definecolor{codepurple}{rgb}{0.58,0,0.82}
\definecolor{backcolour}{rgb}{0.95,0.95,0.92}

\lstdefinestyle{mystyle}{
    backgroundcolor=\color{backcolour},   
    commentstyle=\color{codegreen},
    keywordstyle=\color{magenta},
    numberstyle=\tiny\color{codegray},
    stringstyle=\color{codepurple},
    basicstyle=\ttfamily\footnotesize,
    breakatwhitespace=false,         
    breaklines=true,                 
    captionpos=b,                    
    keepspaces=false,                 
    numbers=left,                    
    numbersep=5pt,                  
    showspaces=false,                
    showstringspaces=false,
    showtabs=false,                  
    tabsize=1
}

\AtBeginDocument{%
  }

\copyrightyear{2026}
\acmYear{2026}
\setcopyright{cc}
\setcctype{by}
\acmConference[DAC '26]{63rd ACM/IEEE Design Automation Conference}{July 26--29, 2026}{Long Beach, CA, USA}
\acmBooktitle{63rd ACM/IEEE Design Automation Conference (DAC '26), July 26--29, 2026, Long Beach, CA, USA}
\acmDOI{10.1145/3770743.3803999}
\acmISBN{979-8-4007-2254-7/2026/07}

\begin{document}

\title{Knowledge-Driven Hybrid SSD Management Enhanced by Fine-Tuned LLMs}

\author{Qian Wei}
\affiliation{%
  \institution{\textit{Shandong University}}
  \city{Qingdao}
  \state{Shandong}
  \country{China}
}

\author{Yi Li}
\affiliation{%
  \institution{\textit{University of Texas at Dallas}}
  \city{Richardson}
  \state{Texas}
  \country{United States}
}

\author{Zehao Chen}
\affiliation{%
  \institution{\textit{Shandong University}}
  \city{Qingdao}
  \state{Shandong}
  \country{China}
}

\author{Tianren Zhou}
\affiliation{%
  \institution{\textit{Shandong University}}
  \city{Qingdao}
  \state{Shandong}
  \country{China}
}

\author{Zhaoyan Shen}
\authornote{Zhaoyan Shen is the corresponding author. Email: shenzhaoyan@sdu.edu.cn.}
\affiliation{%
  \institution{\textit{Shandong University}}
  \city{Qingdao}
  \state{Shandong}
  \country{China}
}

\author{Dongxiao Yu}
\affiliation{%
  \institution{\textit{Shandong University}}
  \city{Qingdao}
  \state{Shandong}
  \country{China}
}

\author{Bingzhe Li}
\affiliation{%
  \institution{\textit{University of Texas at Dallas}}
  \city{Richardson}
  \state{Texas}
  \country{United States}
}

\begin{abstract}
A hybrid Solid-State Drives (SSDs) integrates different modes of flash cells (e.g., single-level cell (SLC) and Quad-Level Cell (QLC)) and enables them to convert between each other, achieving both high performance and storage capacity. However, this hybrid design introduces a significantly larger design space than traditional SSDs with additional design factors such as flash conversion and data migration across different flash modes, leading to higher optimization complexity. Efficient management of such complexity requires deep hybrid SSD knowledge and dynamic adjustment mechanisms. Large language models (LLMs) offer a promising solution through their contextual reasoning and adaptive coordination capabilities.

In this work, we explore the potential of using LLMs in understanding and efficiently managing hybrid SSD design space. We find that leveraging LLMs for knowledge-guided optimization of management parameters enables substantial performance gains. Building on these insights, we propose LLM-hybridSSD, an integrated optimization framework that formulates hybrid SSD management as a parameter-tuning problem, employs an LLM-based tuner for adaptive configuration, and applies reinforcement learning-based fine-tuning to align local lightweight models with domain-specific knowledge. Experimental results show an average 58.92\% increase in throughput and a 28.56\% reduction in write amplification (WA) compared with state-of-the-art schemes under different real-world workloads.
\end{abstract}



\maketitle

\section{Introduction}\label{sec:Introduction}

Hybrid SSDs serve as a versatile storage solution, leveraging the integration of multiple modes of flash memory cells, including SLC and QLC, to simultaneously optimize storage capacity and performance. To achieve dynamic flexibility, hybrid SSDs enable flash cells to switch between different operational modes~\cite{intel, mi}, allowing the system to adapt to diverse application requirements. This mechanism increases the complexity of flash control logic and elevates both computational and memory demands. In response to these challenges, host-managed SSDs expose the physical data layout to the resource-rich host system, providing a more efficient approach to coordinating flash management.

%A key feature of hybrid SSDs is their dynamic flexibility, which enables flash cells to switch between different operational modes~\cite{intel, mi}. 

%%但是混合SSD使得flash的控制逻辑更加复杂，对计算和memory的需求也都更高。面向。。。负载的时候性能波动大。host通过怎么着。。。。提供了一种更好的管理方式 
%The advent of host-managed SSDs, which expose the physical data storage layout to the host system, provides a unique opportunity to leverage hybrid SSDs’ adaptive dynamic mode switching to meet the varying workload requirements.

%The integration of multi-flash modes presents significant management challenges across hardware, algorithmic, and application layers.

%At the algorithmic level, improper timing or granularity of conversions can degrade write performance or lead to underutilized capacity, highlighting the need for adaptive transition control~\cite{intel,mi}.

%aintain efficient data placement for stable performance

% At the algorithmic level, Yoo et al.~\cite{DBLP:conf/hotstorage/YooS20} use machine learning to predict optimal conversion timing and granularity. 
Despite the abundant system resources of host-managed SSDs, the integration of multiple flash modes still presents significant management challenges across both the hardware and application layers. At the hardware level, dynamic flash cell mode conversions must be carefully coordinated with space allocation and garbage collection. Improper timing or granularity in these conversions can lead to performance degradation and underutilized capacity. On the application side, rapidly changing workload patterns alter data access locality, making it difficult for the system to adaptively place data in the most suitable flash modes~\cite{stoica2019understanding}.

To overcome these challenges, Artificial Intelligence (AI) techniques have been increasingly employed to optimize hybrid SSD management. At the hardware level, Yoo et al.~\cite{DBLP:conf/hotstorage/YooS20} and RL-cSSD~\cite{DBLP:conf/dac/WeiLJZSL23} utilize machine learning to optimize mode conversion and garbage collection based on internal device states, aiming to strike a balance between performance and endurance. On the application side, HAML~\cite{DBLP:conf/iccad/LiDYLYD19} leverages kmeans-based hotness classification to guide data placement across different flash modes, effectively enhancing performance stability of hybrid SSD.

%First, existing AI-based schemes rely heavily on algorithm parameters manually configured by domain experts, such as retraining intervals or decision thresholds of HAML, which often require specialized knowledge and extensive trial-and-error, leading to high manual deployment cost and complexity.
However, current AI-driven optimization methods face several inherent limitations. First, many existing AI-based schemes heavily depend on manually designed optimization logic. Domain experts must possess deep knowledge of hybrid SSDs, including aspects such as data placement and mode conversion, and often engage in extensive trial and error to design algorithm parameters. This reliance on expert input leads to high deployment costs and increased complexity. Second, the inherently complex and dynamic nature of hybrid SSDs limits the generalizability of existing AI models. As upper-layer application workloads and internal device states continuously evolve due to shifting access patterns, the lack of dynamic adjustment mechanisms results in inefficient or even incorrect management decisions, ultimately compromising performance and robustness in real-world deployments.

The emergence of LLMs provides new opportunities for hybrid SSD management optimization. Models such as GPT~\cite{DBLP:journals/corr/abs-2303-08774} and LLaMA~\cite{DBLP:journals/corr/abs-2302-13971} possess strong contextual understanding and knowledge transfer ability, enabling rapid adaptation to diverse scenarios with minimal interactions for more efficient optimization. In addition, the reasoning capabilities of LLMs support cross-dimensional decision-making~\cite{DBLP:conf/hotstorage/ThakkarSDS024, DBLP:conf/noms/JeongKNYH24, DBLP:conf/hotstorage/AkewarQMB25}, allowing the generation of globally informed strategies even under interdependent system complexities.
%and planning 

In this paper, we aim to leverage the capabilities of LLMs to enhance the management efficiency of hybrid SSDs. We begin with a preliminary exploration of applying LLMs by directly replacing existing AI-based optimization schemes. The results show that although LLMs demonstrate a basic understanding of system states and management objectives, such direct substitution yields no performance gain while incurring high cost. These findings reveal the need for a more structured integration of LLMs into hybrid SSD management, raising two key challenges: \textbf{(1) How can LLMs be effectively integrated into the appropriate layer of hybrid SSD management?} Using LLMs directly is ineffective without clear guidance on what to adjust and at what granularity, making the definition of LLM-driven optimization scope a key challenge. \textbf{(2) How to integrate certain domain Hybrid SSD knowledge associate with LLM management to preserve both performance and cost efficiency?} Large-scale LLMs deliver strong reasoning capability but are costly to deploy, while lightweight models offer lower overhead but limited understanding of SSD-specific behaviors, making the integration of domain knowledge for efficient and accurate optimization a key challenge.

To address the first challenge, we leverage LLMs to enhance existing hybrid SSD management by formulating the optimization task as a structured parameter-tuning problem with a clearly defined configuration space. Parameters are categorized into SSD-specific, workload-related, and AI strategy-related groups, and performance-sensitive parameters are selected to reduce tuning complexity while maximizing optimization effect. Based on the formulation, we design an LLM-based tuner that transforms normalized workload and device-state information into structured prompts and generates configuration recommendations. The generated outputs are validated through constraint checking to ensure correctness.

%To handle the second challenge, Large LLMs provide strong reasoning ability but are costly to deploy, while lightweight local models often lack sufficient understanding of SSD-specific behaviors. 
% deploy a lightweight local model and fine-tune it to ensure competitive performance while maintaining low cost.
To address the second challenge, we integrate local lightweight LLM deployment with fine-tuning techniques to achieve competitive management performance at low cost. We employ a Guided Reinforcement Policy Optimization (GRPO) mechanism that enables the lightweight model to continuously learn from runtime feedback and progressively internalize hybrid SSD-specific knowledge. This process produces domain-specialized LLMs that not only understand hybrid SSD behaviors more accurately but also retain low computational and storage footprint.

%To address the second challenge, we integrate on-device lightweight LLM deployment with fine-tuning techniques to achieve competitive management performance while maintaining low cost. we employ a Guided Reinforcement Policy Optimization (GRPO) mechanism that enables the lightweight model to continuously learn from runtime feedback and align its behavior with hybrid SSD characteristics. This fine-tuning builds domain-specialized LLMs capable of efficient configuration adaptation while maintaining low computational and storage overhead.

%Our design adopts a Guided Reinforcement Policy Optimization (GRPO) mechanism that fine-tunes a local lightweight model to build domain-specialized versions tailored for hybrid SSD management, enabling efficient and domain-aware optimization.

%To address this issue, we propose fine-tuning local lightweight models to build domain-specialized versions tailored for hybrid SSD management. Specifically, we adopt a Guided Reinforcement Policy Optimization (GRPO) strategy, which leverages the observed impact of different system settings to guide model parameter updates.

Building on the above designs, we develop an integrated optimization framework, named LLM-hybridSSD. The framework integrates task formulation, configuration tuning, and lightweight model adaptation into a unified LLM-based feedback loop. Experimental results show that LLM-hybridSSD significantly enhances performance, increasing throughput by 58.92\% while reducing WA by 28.56\% under real-world traces compared with SOTA schemes.

%Experimental results show that LLM-hybridSSD significantly enhances device performance, increasing throughput by 58.92\% and WA by 28.56\% under real-world workloads compared with SOTA schemes.

%A LLM-based tuner analyzes performance history and device states to adapt system parameters dynamically, while a validation module ensures stability and correctness.

The contributions of this work are summarized as follows:
\begin{itemize}
    \item We formalize hybrid SSD management as a structured parameter tuning problem, introducing a well-defined configuration space that categorizes SSD-specific, workload-related and AI strategy-related parameters while identifying performance critical factors to reduce tuning complexity.
    \item We design and fine-tune lightweight LLMs tailored for cost-constrained hybrid SSD environments, enabling efficient deployment and continuous learning through a Guided Reinforcement Policy Optimization mechanism.
    \item We develop LLM-hybridSSD, an integrated optimization framework that unifies task formulation, configuration tuning, and local lightweight model adaptation into a closed-loop LLM-enhanced hybrid SSD management system.
    %The framework is open-sourced\footnote{https://anonymous.4open.science/r/DAC-609.}.
    \item We conduct comprehensive evaluations under real-world workloads, demonstrating that LLM-hybridSSD substantially outperforms state-of-the-art schemes.
\end{itemize}

The remainder of this paper is organized as follows: \S~\ref{sec:background} provides the background. \S~\ref{sec:moti} explores the potential of LLMs in hybrid SSD management. \S~\ref{sec:design} presents the LLM-hybridSSD framework, and \S~\ref{sec:results} discusses experimental results. Finally, \S~\ref{sec:conclusion} concludes this paper.

\section{Background}\label{sec:background}

\subsection{Hybrid SSDs\label{2.2}}

\begin{figure}[t]
    \centering
    \includegraphics[width = 3.3in]{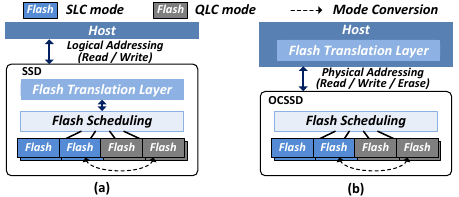}
     \vspace{-0.1cm}
    \caption{The core module of (a) traditional SSD (b) host-managed SSD.}
    \label{ocssd}
    \vspace{-0.2cm}
\end{figure}

A flash-based SSD consists of multiple flash memory chips and a controller that handles address mapping, wear leveling, and garbage collection through the Flash Translation Layer (FTL)~\cite{DBLP:conf/dac/WeiLJZSL23, DBLP:conf/isca/HuangBQS15, DBLP:conf/fast/ZhuSWCYYS25}, as illustrated in Figure~\ref{ocssd}. To improve flexibility and reduce device-side complexity, host-managed SSDs shift FTL management responsibilities to the host operating system~\cite{DBLP:journals/tos/KwakLPJS20}.

Flash memory stores multiple bits per cell through electron injection, enabling configurations such as SLC and QLC. While higher-density technologies like QLC provide substantially increased storage capacity, they suffer from lower endurance and degraded performance. Hybrid SSDs integrate multiple flash modes within a single device (e.g., SLC and QLC), and their cells can be reconfigured between modes~\cite{stoica2019understanding, marotta2013dynamic}, allowing the system to flexibly trade off between performance and capacity. However, managing such hybrid architectures is challenging because data must be dynamically placed and migrated across regions to sustain both efficiency and device longevity.

%The SSD controller manages different cell types through shared channels to achieve optimized capacity and performance simultaneously. 
%Different types of flash cells are convertible by changing the threshold voltages and Incremental Step Pulse Program.
%in hybrid SSDs

%To manage data placement and mode conversion in hybrid SSDs, various optimization mechanisms have been explored.

AI-based optimization schemes have been proposed for hybrid SSD management, aiming to model workload access behaviors and improve data placement and migration. HAML~\cite{DBLP:conf/iccad/LiDYLYD19} exemplifies this direction by using k-means clustering on update frequency and access intervals to group data with similar hotness, reducing interference from varying workloads. From another perspective, RL-based schemes target adaptive control of flash-mode conversion. RL-cSSD~\cite{DBLP:conf/dac/WeiLJZSL23} leverages environment feedback to refine conversion timing and alleviate performance degradation caused by inefficient mode conversions.

\subsection{Large Language Models}

Large language Models (LLMs), such as GPT~\cite{DBLP:journals/corr/abs-2303-08774}, Meta LLaMA~\cite{DBLP:journals/corr/abs-2302-13971} and Deepseek~\cite{DBLP:journals/corr/abs-2401-02954}, demonstrate strong reasoning and generalization capabilities, enabling their application beyond language tasks. 
By modeling complex relationships and high-dimensional objectives~\cite{hornik1991approximation}, LLMs offer new opportunities for intelligent configuration and adaptive management of large-scale computing systems.

LLMs are typically applied to system and application optimization through two deployment paradigms: cloud-based and local deployment~\cite{DBLP:journals/csur/ZhengCQSSC25, DBLP:conf/appt/WeiCZZJZS25}. Cloud-based solutions leverage centralized resources and offer convenient access, but incur recurring costs, raise potential privacy concerns, and often rely on closed-source models. Local deployment enhances data security, avoids network latency, and can utilize open-source models for greater control, but requires substantial hardware resources and user-managed maintenance.

Fine-tuning has become a central technique for adapting LLMs to downstream tasks. Full-parameter fine-tuning~\cite{DBLP:conf/nips/BrownMRSKDNSSAA20} offers strong accuracy but incurs high computational cost, while parameter-efficient methods such as adapters~\cite{DBLP:conf/icml/HoulsbyGJMLGAG19}, prefix-tuning~\cite{DBLP:conf/acl/LiL20}, and LoRA~\cite{DBLP:conf/iclr/HuSWALWWC22} adjust only small subsets of weights for efficiency. Further, reinforcement learning-based tuning~\cite{DBLP:journals/corr/abs-2009-01325, openai2024grpo} further aligns model behavior with desired objectives through feedback-driven optimization.

\begin{figure}[t]
    \centering
    \small
    \begin{tabular}{cc}
        \hspace{-4mm}
        \includegraphics[width=0.25\textwidth]{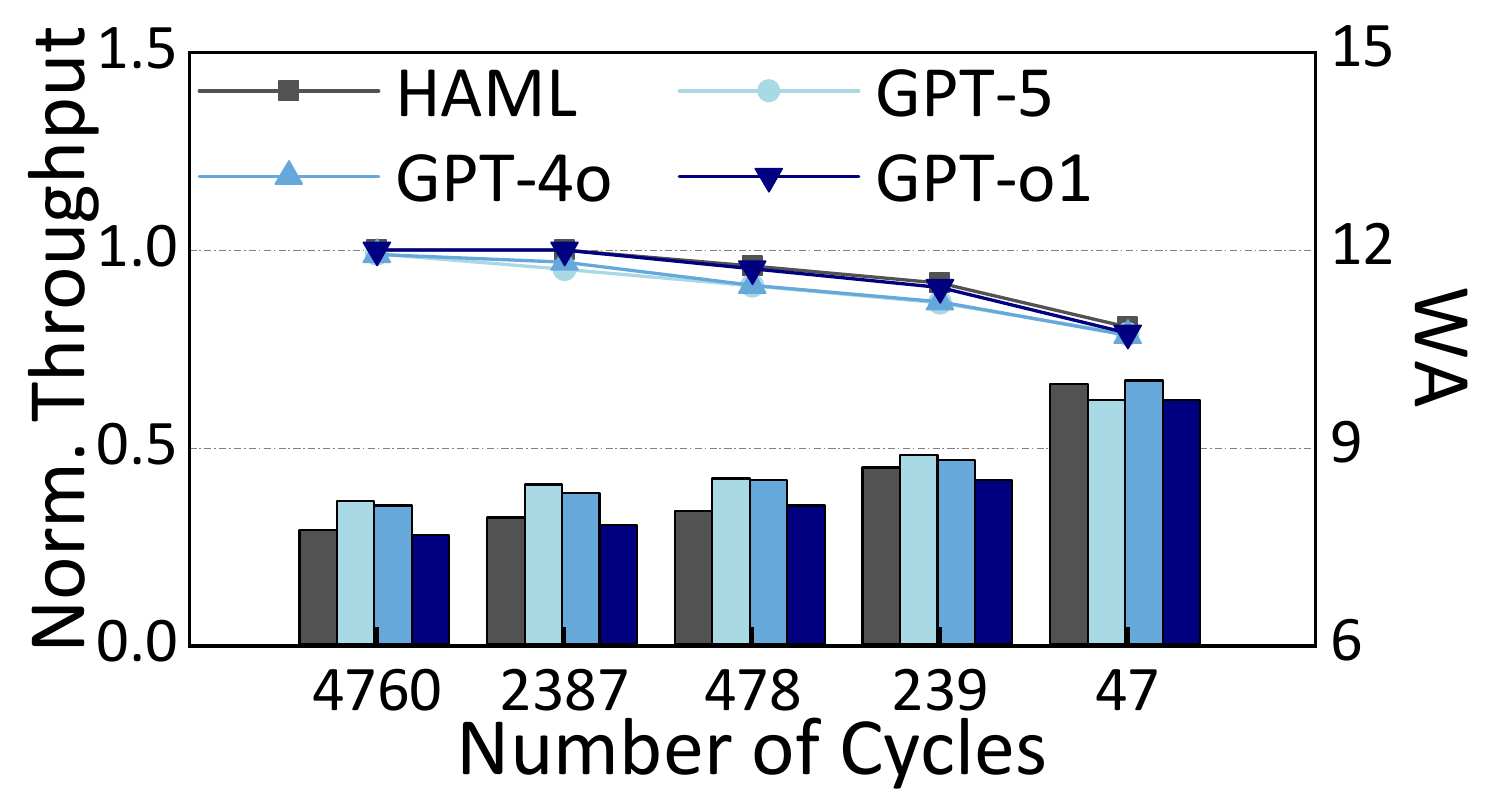} &
        \hspace{-4mm}
        \includegraphics[width=0.257\textwidth]{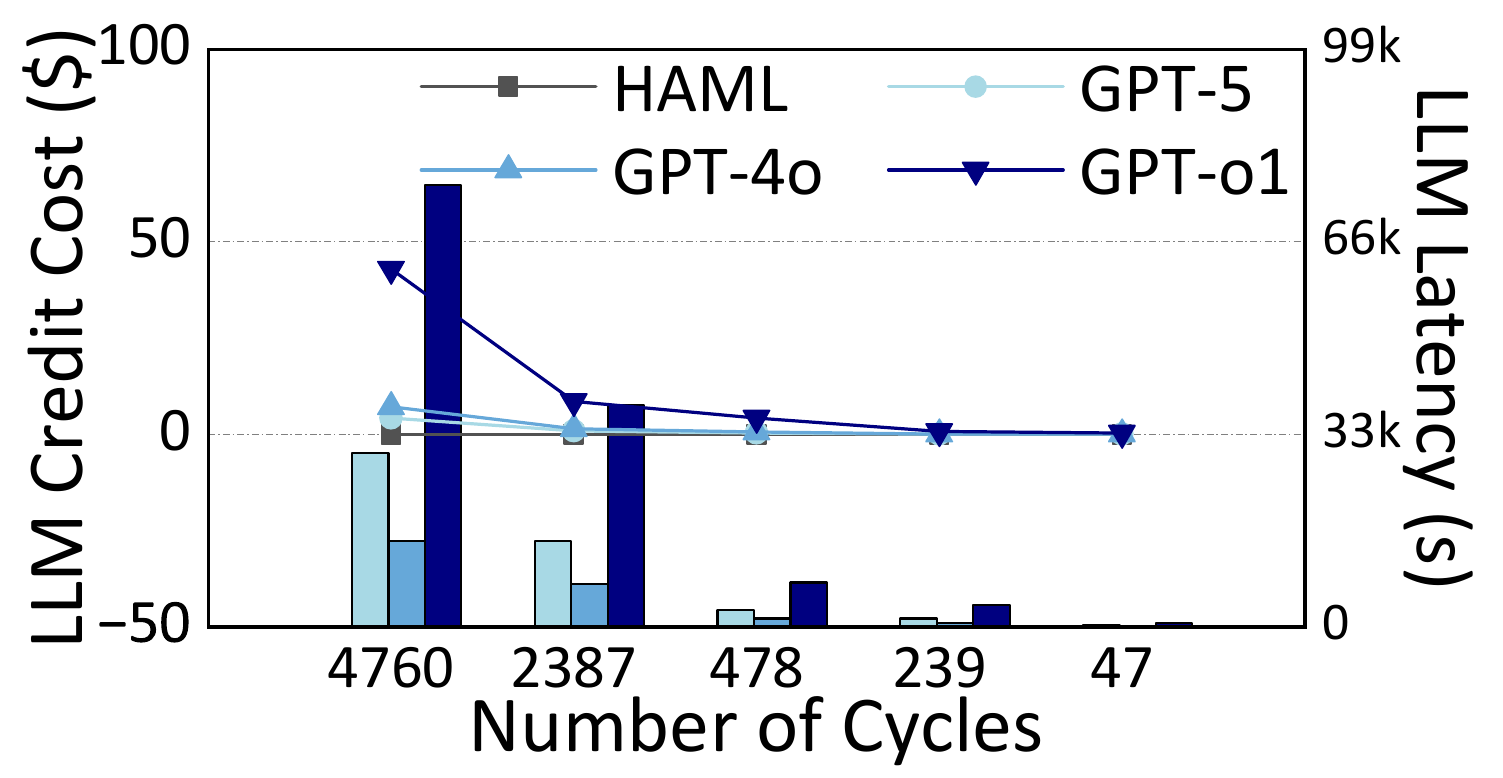} \\
        \hspace{-5mm}(a) Performance. & \hspace{-4mm}(b) LLM Overhead. \\
    \end{tabular}
    \vspace{-0.1cm}
    \caption{\label{pic:MOTI1}Performance (left axis, lines) and cost (right axis, bars) with different reclassification cycles.}
    \vspace{-0.2cm}
\end{figure}
\section{Optimization Potential of LLMs}\label{sec:moti}

In this section, we conduct preliminary experiments to examine the potential of LLMs in hybrid SSD management, focusing on whether they can replace existing AI-based approaches or further enhance them through parameter tuning. The evaluation setup follows the configuration described in \S~\ref{sec:results}.
%Our exploration 

\subsection{Handling Primary Management by LLMs}\label{sec:moti1}

\begin{figure}[t]
    \centering
    \small
    \begin{tabular}{cc}
        \hspace{-4mm}
        \includegraphics[width=0.249\textwidth]{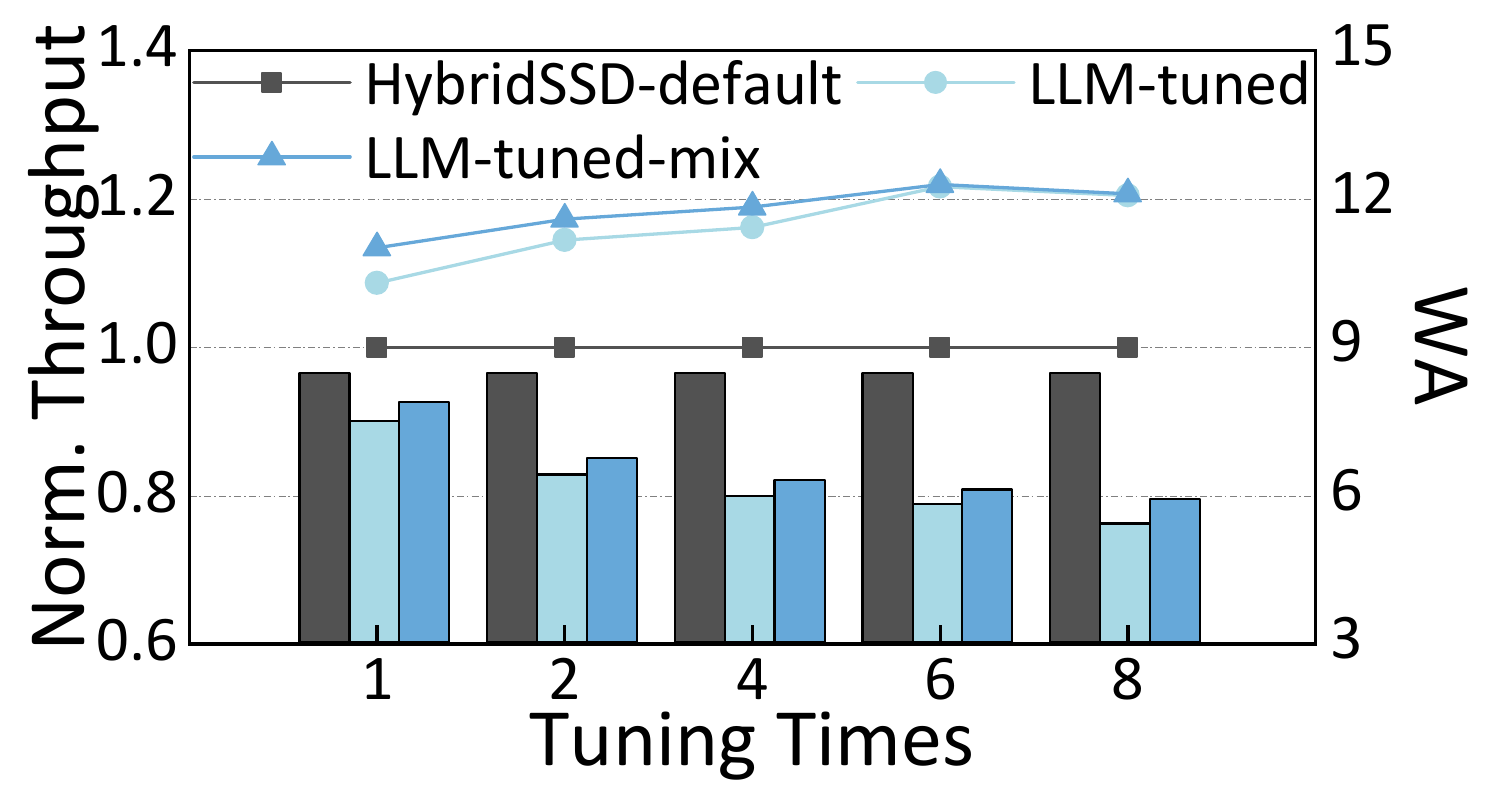} &\hspace{-3.5mm}
        \includegraphics[width=0.257\textwidth]{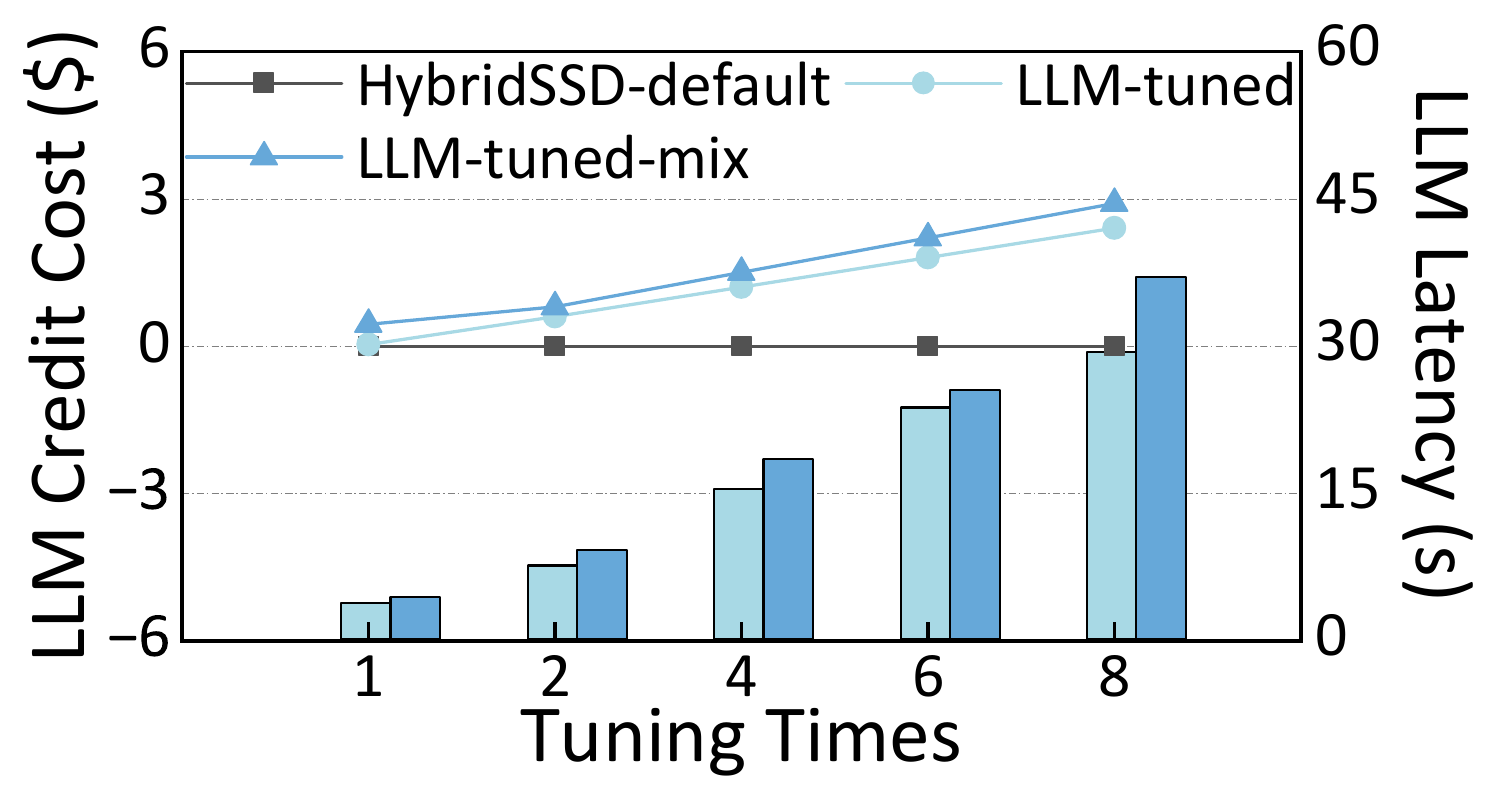} \\
        \hspace{-5mm}(a) Norm. Exec. Time. &\hspace{-4mm} (b) LLM Overhead. \\
    \end{tabular}
    \vspace{-0.1cm}
    \caption{\label{pic:MOTI3}Performance (left axis, lines) and cost (right axis, bars) of different tuning schemes.}
     \vspace{-0.2cm}
\end{figure}

%Using the prn\_0 workload~\cite{narayanan2008write}, we provide workload information to LLMs, requesting an analysis of the data hotness classification file for each calling cycle as input for subsequent cycle.

To evaluate whether LLMs can handle fundamental hybrid SSD management tasks, we replace the ML-based data hotness classification scheme (HAML~\cite{DBLP:conf/iccad/LiDYLYD19})  with advanced large-scale LLMs (i.e., GPT-5, GPT-4o, and GPT-o1) via API calls on the host-managed SSD. We serve the trace data (i.e., prn\_0~\cite{narayanan2008write}) to each model and request it to generate a data hotness classification file, which is then used to guide subsequent data placement decisions.

Figure~\ref{pic:MOTI1} shows the results with different reclassification cycles. As shown in Figure~\ref{pic:MOTI1} (a), under frequent cycles, LLM-based schemes provide similar performance to ML-based schemes, with normalized throughput fluctuation within 2.3\%. This reveals that LLM can effectively handle the hotness classification task. However, reducing the reclassification frequency significantly degrades performance for both ML-based and LLM-based schemes, since infrequent hotness-classification updates fail to capture rapid workload changes. Meanwhile, as shown in Figure~\ref{pic:MOTI1} (b), although the performance is comparable, the LLM-based approach incurs notable credit cost (up to \$42.97) and latency (up to \textasciitilde20 hrs) overhead, making direct deployment impractical.

In summary, large-scale LLMs are trained with hybrid SSD knowledge and achieve as good performance as traditional schemes. However, directly applying LLMs to hybrid SSD management (e.g., simply replacing existing schemes) does not yield significant performance gains while introducing unacceptable cost.

\subsection{Transferring SSD Tuning into LLMs}\label{3.3}

Since directly replacing existing schemes with LLMs does not yield substantial performance improvement, we instead investigate how LLMs can be used to assist existing algorithms. Specifically, we explore whether LLMs can enhance current ML-based strategies by tuning their key parameters. Hybrid SSD management involves a vast and complex parameter space covering numerous configurable aspects. These parameters, such as garbage collection thresholds and data migration frequency, work together to guide system behavior and ensure stable and efficient operation.

%Building on the identified optimization potential,

To ground our study, we select HAML, a state-of-the-art ML-based design, as the baseline method. we then compare three configurations: \textit{HAML}, which uses static parameters; \textit{LLM-tuned}, where best-performing GPT-o1 dynamically adjusts HAML-related parameters such as retraining interval and K-means thresholds; \textit{LLM-tuned-mix}, which expands the tuning space to include additional GC/MC granularity parameters.

%Since directly replacing existing schemes with LLMs does not yield substantial performance improvement, we instead investigate how LLMs can be used to assist existing algorithms. Specifically, we explore whether LLMs can enhance current ML-based strategies by tuning their key parameters. Hybrid SSD management involves a large and complex parameter space—such as garbage collection thresholds and data migration frequency—whose interactions shape system behavior and influence overall system efficiency.

%To ground our study, we select HAML, a state-of-the-art ML-based design, as the baseline method. We then compare three configurations: HAML, which relies on static, manually chosen parameters; LLM-tuned, in which the best-performing GPT-o1 dynamically adjusts HAML-related parameters such as retraining intervals and k-means clustering thresholds; and LLM-tuned-mix, which further broadens the tuning space by incorporating additional GC/MC granularity parameters.

%These parameters, including garbage-collection thresholds, data migration frequency, and caching ratios, collectively determine the overall trade-off among performance, endurance, and efficiency.

%we further investigate whether LLMs can effectively perform parameter tuning for hybrid SSD management.

%; and \textit{LLaMA-tuned-mix}, which directly applies a locally deployed lightweight model to conduct parameter adjustments

Figure~\ref{pic:MOTI3} shows the  performance and cost of these three scheme with different tuning times under the prn\_0 workload~\cite{narayanan2008write}. As shown, the LLM-tuned configurations achieve noticeable improvements over HAML. Specifically, LLM-tuned configuration improves throughput by 16.28\% and decreases WA by 15.2\%, indicating that LLMs effectively generate parameter adjustments. When the parameter space expands in LLM-tuned-mix, performance further improves, with throughput increasing by 20.47\% and WA dropping by 18.7\% compared with HAML. This result demonstrates that LLMs can adapt flexibly to an expanding parameter pool, refining configurations and enhancing system performance accordingly. However, the tuning process also incurs additional overhead due to repeated LLM invocations. In our experiments, the LLM-based approach incurs credit cost (up to \$2.92) and latency (up to \textasciitilde36.9 s) overhead, partially offsetting the performance gains.

%However, the locally deployed LLaMA 3.2 model, though eliminating API costs, achieves smaller gains due to its limited parameter capacity. This limitation reduces its ability to capture diverse workload patterns and the fine-grained correlations required for precise configuration optimization, resulting in performance that is 9.2\% to 21.4\% lower than that of GPT-4-turbo.

%In summary, LLMs show strong potential for managing complex parameter-tuning scenarios in hybrid SSDs. 

In summary, LLMs show strong potential to optimize hybrid SSD management through dynamic parameter-tuning strategies. However, realizing their full potential requires a well-structured tuning framework that balances optimization efficiency and cost.

%Lightweight models such as LLaMA 3.2 offer efficient and low-cost deployment, yet their limited capacity restricts their ability to capture diverse workload characteristics, leaving a persistent performance gap compared with large-scale models.

%\input{4_para}

\section{System Design}\label{sec:design}

\begin{figure}[t]
    \centering
    \includegraphics[width = 3.1in]{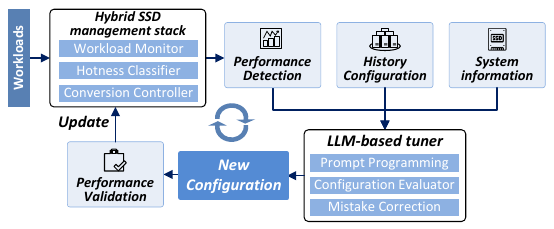}
     \vspace{-0.1cm}
    \caption{The overall framework of LLM-hybridSSD.}
    \label{Architecture}
     \vspace{-0.4cm}
\end{figure}

\subsection{System Overview}

%In LLM-hybridSSD, we design an optimization framework that employs a locally fine-tuned LLM to perform configuration tuning for more efficient hybrid SSD management. The fine-tuned model is incorporated into the management system to enable adaptive and coordinated control across modules, as illustrated in Figure~\ref{Architecture}.

In this work, we propose LLM-hybridSSD, a knowledge-driven hybrid SSD management framework that integrates an LLM-based tuner into the SSD management stack to enable adaptive parameter tuning under dynamic workloads, as shown in Figure~\ref{Architecture}.

% For workload monitor, we design a sliding window-based mechanism that tracks access patterns and detects variations through standard deviation analysis. The hotness classifier follows the HAML scheme~\cite{DBLP:conf/iccad/LiDYLYD19}, while the conversion control adopts the RL-cSSD strategy~\cite{DBLP:conf/dac/WeiLJZSL23}. 
The framework consists of three main components. The \emph{hybrid SSD management stack} provides the core management control for hybrid SSDs. Within this stack, we design a sliding window-based mechanism that tracks access patterns and detects variations to support management decisions. The \emph{LLM-based tuner} employs a fine-tuned local deployed LLM to adjust the carefully designed hybrid SSD configuration space and produce parameter updates. The \emph{performance validation module} supplies feedback to the tuning loop based on short-term system behavior.

The framework operates as a closed feedback loop. At each tuning interval, runtime statistics are collected and encoded as inputs to the LLM-based tuner. The tuner produces a candidate configuration that is applied for a short evaluation window. The performance validation module then measures the corresponding execution metrics and compares them with those from the previous interval. If the candidate configuration satisfies the acceptance criterion, it is adopted as the new management policy. Otherwise, the system restores the previous configuration and may request additional candidates.

\subsection{Problem Formulate}\label{sec:4.2}

%\subsubsection{Parameter Options of Hybrid SSDs\label{sec4.1}}

In this work, we do not modify the SSD hardware design itself, in contrast to AutoBox~\cite{DBLP:conf/micro/LiS023} which targets hardware configuration during device development. Instead, given a fixed hardware design, we focus on the runtime control policy and formulate it as a structured parameter-tuning problem. This formulation identifies the dynamically adjustable software parameters that meaningfully influence system behavior, enabling the LLM to operate within a well-defined configuration space rather than over raw device states. We group the tunable parameters into three main categories:

$\bullet${ SSD-specific parameters:} These parameters are inherent to SSD firmware, controlling flash memory allocation and conversion. They are crucial for optimizing performance and enhancing device stability. Typical examples include GC threshold and granularity.

$\bullet${ Workload-related parameters:} These parameters are used to collect data access patterns and characteristics, which help hybrid SSD management policies better understand current workloads. A typical example is the workload sliding window size.

$\bullet${AI Strategy-related parameters:} These parameters are key to AI-driven algorithms. Specifically, they tune the behavior of these strategies, such as ML-based hotness classification and RL-based space management, ensuring better performance and reliability. Examples include K-means max iterations and RL training interval.

Initially, we consider a broad set of parameters spanning these categories. However, including all parameters can be counterproductive, as an large parameter space may exceed the context window of LLM and tuning irrelevant parameters may introduce noise, reducing convergence efficiency and stability~\cite{DBLP:conf/hotstorage/ThakkarSDS024}. Therefore, we focus on identifying the most performance-sensitive parameters.

\begin{table}[]
\centering
\caption{Sensitive Parameters with Default Values}\label{tab:paras}
\vspace{-0.2cm}
\resizebox{0.47\textwidth}{!}{
\begin{tabular}{ccc}
\hline
\multicolumn{1}{c}{\textbf{Category}} &
\multicolumn{1}{c}{\textbf{Parameter Name}} &
\multicolumn{1}{c}{\textbf{Default Value}} \\ \hline
\multirow{5}{*}{SSD-specific}     
& Conversion granularity       & 1      \\
& Conversion trigger threshold & 6      \\
& GC granularity               & 1      \\
& GC trigger threshold         & 6      \\
& Data placement strategy      & SLC first \\ \hline
\multirow{3}{*}{Workload-related} 
& Window size                  & 2000   \\
& Standard deviation threshold & 10000  \\
& Slice size                   & 200MB  \\ \hline
\multirow{7}{*}{AI Strategy-related} 
& K-means max iterations       & 10     \\
& K-means trigger threshold    & 10000  \\
& RL training interval         & 1000   \\
& RL learning rate             & 0.1    \\
& RL reward                    & 1.6ms  \\
& RL discount factor           & 0.9    \\
& RL exploration rate          & 0.1    \\ \hline
\end{tabular}}
\end{table}

Through offline workload analysis, we evaluate the impact of each parameter across different workload scenarios. Parameters with minimal influence, such as the K-means convergence tolerance (tol) whose variations seldom affect clustering outcomes or overall performance, are excluded from consideration. Based on this sensitivity analysis, we select 15 key parameters from an initial set of 32 across the three categories. The selected parameters and their default values, derived from existing strategy settings, are summarized in Table~\ref{tab:paras}.

\subsection{LLM-based Tuner of SSD Configurations\label{sec:4.3}}

%Our solution allows the tuner to support both API calls and local deployment from host. This flexibility enables users to choose the most suitable approach based on their requirements, with a discussion on performance and overhead provided in Subsection~\ref{6.3}.

This subsection mainly introduces the working flow of LLM-based tuner. There are three major phases: 1) Prompt Engineering, 2) Configuration Generation, and 3) Mistake Correction.

%The optimization targets focus on two key metrics: throughput and WA.
\textbf{Prompt Engineering.} The prompt engineering phase is used to provide detailed and accurate background information for LLM to generate effective responses. The prompt is organized into five key stages. First, \ding{182}\textbf{Role Assignment} is carried out by designating the LLM as an SSD expert. Then, during \ding{183}\textbf{Hybrid SSD Overview} and \ding{184}\textbf{SSD Management}, the LLM is provided with system information, which is categorized into the inherent configuration and the current optimization strategies. In \ding{185}\textbf{Historical Configs \& Performance}, a detailed list of past configuration changes, their reasons, and performance impacts is included. Finally, \ding{186}\textbf{Specify Requirements} ensures the LLM generates a new configuration list with improved accuracy by defining a clear output format. We give an example of the prompt and its corresponding output:
\begin{shaded}
\textbf{\textit{Input: }}\ding{182} You are an SSD Expert. You are being consulted to improve the SSD configuration by optimizing options file based on system information and benchmark output. \ding{183} The SSD consists of SLC and QLC modes... $<$SSD Settings$>$. \ding{184} The current SSD scheme is as follows: (1) The hotness classification for is using the K-means algorithm… ; (2) A space management scheme based on Q-learning is employed… ; …; \ding{185} First, the historical performance and configuration changes are shown. The point in time configuration changes are as follows: …, The performance is: … The current option file is: … \ding{186} Based on these information generate a new file in the same format as the options\_file to improve the SSD throughput and WA. Enclose the new options file in ` `.

\textbf{\textit{Output: }}New configuration:

`1. K-means trigger threshold: 1000; 2.Windows size: 1500; ...`

Reason: …
\end{shaded}

The input prompt size typically remains within the context limits of mainstream LLMs. To further ensure safety under iterative adjustments, we adopt a sliding window approach, splitting long texts into overlapping segments while retaining the ending of the previous segment to maintain context and avoid information loss.

%We estimate the final prompt length to be around 2500 tokens after 10 adjustments per trace, well within the 4096-token limit of LLaMA 3.2(通过什么样的方式可以控制output token的长度，满足不同模型的限制内). Further, to prevent token overflow from iterative adjustments, we adopt a sliding window approach, splitting long texts into overlapping segments while retaining the ending of the previous segment to maintain context and avoid information loss.

\textbf{Configuration Generation.} The configuration generation phase is used to transform the natural language output of LLM into a machine-readable format. This module extracts two key components: the reason for the adjustment and the updated configuration list. The historical updates are recorded for future tuning reference.

\textbf{Mistake Correction.} The mistake correction phase validates the updated configuration through format checking and numerical range checking. For format checking, we verify that each configuration matches the correct format and remove any irrelevant descriptions. For numerical range checking, parameter thresholds are set to maintain the correctness of the ranges, preventing performance degradation due to excessive tuning.

%This phase validates the updated configuration through format checking and numerical range checking. Format checking ensures each entry follows the required structure, and range checking enforces valid parameter bounds to prevent errors or unstable tuning.

\subsection{RL-driven Fine-Tuning of Lightweight LLMs}\label{sec:4.4}

\begin{figure}[t]
    \centering
    \includegraphics[width = 3.3in]{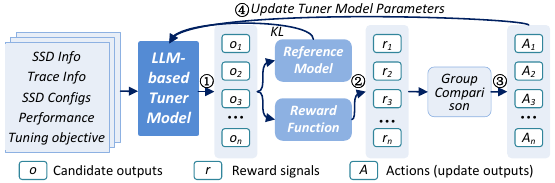}
    \caption{Fine-Tuning workflow under the GRPO strategy.}
    \label{fine_tuned}
     \vspace{-0.2cm}
\end{figure}

As discussed in \S~\ref{3.3}, deploying large-scale LLMs for hybrid SSD management introduces excessive cost, making them unsuitable for real-world deployment. To address this, we adopt local lightweight models that can operate entirely on a single machine. However, their limited capacity results in insufficient hybrid SSD domain knowledge, causing inaccurate decisions and performance deviation. To address this issue, we fine-tune the lightweight model to help it better understand hybrid SSD domain knowledge and effectively handle management tasks. Specifically, we adopt the Guided Reinforcement Policy Optimization (GRPO) strategy~\cite{openai2024grpo} to perform this fine-tuning process.

%To overcome this limitation, we employ reinforcement learning (RL)-based fine-tuning, which allows the model to learn from runtime feedback and continuously refine configuration strategies. In particular, we adopt the Guided Reinforcement Policy Optimization (GRPO) strategy, which improves stability and convergence through group-wise comparisons.

%In hybrid SSD management, lightweight models often suffer from limited parameter capacity, making it difficult to capture the highly dynamic workload characteristics and diverse device settings. Static fine-tuning alone cannot adequately address this issue, since it lacks adaptability to runtime feedback and struggles to generalize across heterogeneous traces. 

%To overcome these limitations, we employ reinforcement learning (RL)-based fine-tuning, which enables the model to interact directly with the SSD environment, evaluate different configuration schemes, and iteratively improve its decision-making ability. In particular, we adopt the Guided Reinforcement Policy Optimization (GRPO) strategy, which enhances conventional reinforcement learning by leveraging group-wise comparisons and has been recently adopted in advanced LLM alignment.

The fine-tuning workflow is illustrated in Figure~\ref{fine_tuned}. \ding{172} The LLM-based tuner proposed in \S~\ref{sec:4.3} receives system information, including SSD info, trace info, configurations and tuning objectives, and generates multiple candidate configuration files (o1, …, on) representing potential configuration adjustments. \ding{173} Each candidate file is executed within a short evaluation to obtain real SSD performance feedback. Based on the results, a frozen reference model and a reward function compute corresponding reward signals (r1, …, rn). \ding{174} GRPO performs group-wise comparison to identify the candidates with the highest rewards and generates new performance-friendly configuration updates (A1, A2, …, An). \ding{175} The reward signals and the differences among high-reward candidates are then used to update the tuner model parameters, completing the optimization loop. Through this iterative process, the tuner model internalizes hybrid SSD domain knowledge and applies it to balance performance goals and resource constraints, achieving adaptive and stable management under diverse workload and device conditions.

%As shown in Figure~\ref{fine_tuned}, the fine-tuning process begins with the LLM-guided tuner model, which receives related information, including device configuration, workload traces, performance metrics, and tuning objectives.Based on the input, the tuner generates multiple candidate outputs (o1, …, on), each representing a potential parameters adjustment. These candidates are simultaneously evaluated by a frozen reference model and a reward function according to performance feedback, producing a set of reward signals (r1, …, rn). Subsequently, the candidates and their corresponding rewards are aggregated in a group computation step, where the relative effectiveness of different adjustment schemes is compared. From this stage, a set of actions (A1, A2, …, An) is derived, corresponding to executable SSD configuration adjustments. The reward signals are further used to update the tuner model parameters, closing the reinforcement learning loop. The final step applies SSD configuration updates only if the measured performance metrics demonstrate improvement. Through this iterative process, the tuner model progressively learns to balance performance objectives and resource constraints, enabling effective adaptation to diverse workloads and device settings.

\section{Experimental Results}\label{sec:results}

%To evaluate the performance of our methodology, we address four research questions in our experiments:

%In this section, we answer the following questions:

%(1) How effective of the proposed LLM-hybridSSD compared to existing technologies? (\S~\ref{sec6.2})
%(Subsection~\ref{sec6.2})

%(2) How do performance and overhead compare between \textbf{\textit{Fine-tuned local deployment}} versus \textbf{\textit{API calls}} of different LLMs? (\S~\ref{6.3})

%(3) How sensitive is the performance improvement to changes in the environment/settings? (\S~\ref{sec6.5})

\subsection{Environment Setup}

\begin{table}[]
\centering
\caption{SSD Configurations}\label{tab:ssd}
\vspace{-0.2cm}
\resizebox{0.47\textwidth}{!}{
\begin{tabular}{ccccc}
\hline
\multicolumn{1}{c}{\textbf{Parameter}} & 
\multicolumn{1}{c}{\textbf{Value}} & 
\multicolumn{1}{c}{\textbf{Parameter}} & 
\multicolumn{1}{c}{\textbf{Value}} \\ \hline
Capacity                & 256GB  & Write latency (SLC mode) & 200us \\
\#Channels              & 32     & Write latency (QLC mode) & 2ms   \\
Page size               & 16KB   & Read latency (SLC mode)  & 20us  \\
Over-provisioning ratio & 12.5\% & Read latency (QLC mode)  & 140us \\
Pages/block (SLC mode)  & 256    & Erase latency (SLC mode) & 3ms   \\
Pages/block (QLC mode)  & 1024   & Erase latency (QLC mode) & 3.5ms \\ \hline
\end{tabular}}
\end{table}

\begin{table}[]
\centering
\caption{Workload Characteristics}\label{tab:workload}
 \vspace{-0.2cm}
%\footnotesize
\resizebox{0.47\textwidth}{!}{
\begin{tabular}{ccccc}
\hline
\multicolumn{1}{c}{\textbf{Category}} & 
\multicolumn{1}{c}{\textbf{Workloads}} & 
\multicolumn{1}{c}{\textbf{Read request (GB)}} & 
\multicolumn{1}{c}{\textbf{Write request (GB)}} \\ \hline
 & prn\_0             & 13.12      &  45.96                      \\
\multirow{-2}{*}{MSR~\cite{narayanan2008write}}  & usr\_0      & 35.3     & 13.08             \\
FIU~\cite{snia-trace-block-io-388}     & homes         &   5.89              &  39.6           \\
OLTP~\cite{spcspc}     & Financial      &   2.65           &  14.57        \\
KV store~\cite{DBLP:journals/tos/YadgarGJS21}    & ssdtrace       &  199.36      &  46.6        \\
Mix    & mixed\_trace       &  58.38      & 79.52        \\ \hline
\end{tabular}}
\end{table}

\textbf{Hardware Platform.} We evaluate LLM-hybridSSD using a trace-driven hybrid SSD simulator based on SSDsim~\cite{hu2013exploring}. The simulator models hybrid architectures, incorporating SLC-QLC mode conversion, data allocation, and garbage collection. The SSD is configured with a 256 GB capacity and a 16 KB page size, with detailed setups provided in Table~\ref{tab:ssd}. To ensure comprehensive coverage of all possible scenarios and to expedite the testing process, the available SSD space is limited to 32 GB during the experiments. LLM-hybridSSD is executed on a server equipped with 48 Intel Xeon CPUs operating at 2.1 GHz, where both SSD management and the LLM-based tuner run on the host CPU, adhering to the host-managed SSD paradigm.

\textbf{Traces and Baselines.} We use 5 different workloads (shown in Table~\ref{tab:workload}), which cover various scenarios including KV stores, research, financial, etc. Furthermore, to simulate more complex scenarios involving multiple applications and frequent load variations, we construct a mixed trace, denoted as "Mix", by combining several individual traces (i.e., prn\_0, usr\_0, and hm\_0). We compare the proposed LLM-hybridSSD with three other hybrid SSD designs: HAML~\cite{DBLP:conf/iccad/LiDYLYD19}, RL-cSSD (RL-hybridSSD)~\cite{DBLP:conf/dac/WeiLJZSL23} and LLM-hybridSSD (default). HAML employs ML-assisted methods to cluster data with similar hotness. RL-hybridSSD designs an RL-assisted space management scheme to coordinate GC and MC processes. LLM-hybridSSD (default) employs the same optimization strategy as LLM-hybridSSD but uses fixed default configurations during the entire workload execution.

\textbf{LLM Models.} For LLM model setup, we conduct GRPO-based fine-tuning on local deployed LLaMA 3.2, which is a widely adopted lightweight model. During the fine-tuning process, we choose the traces (i.e. mds\_0, prxy\_0, web\_0 form MSR~\cite{narayanan2008write}, mail form FIU~\cite{snia-trace-block-io-388}, storage form Baleen~\cite{DBLP:conf/fast/WongWMGLKSBBG24}) as the training dataset. Additionally, we implement other LLM-driven prototypes using models such as GPT-o1, DeepSeek-R1, and Qwen3, and compare their performance and cost with that of our fine-tuned model (in \S~\ref{5.4} and \S~\ref{sec:5.5}).

%In the evaluation, workloads with different access patterns in MSR~\cite{narayanan2008write}, FIU~\cite{snia-trace-block-io-388}, OLTP~\cite{spcspc} and SSD traces~\cite{DBLP:journals/tos/YadgarGJS21} are selected. Table~\ref{tab:workload} shows the characteristics of these traces. 

\subsection{Overall Performance\label{sec6.2}}

\begin{figure}[t]
	\centering
	\small
	\begin{tabular}{cc}
		\hspace{-5mm}	
        \includegraphics[width=0.24\textwidth]{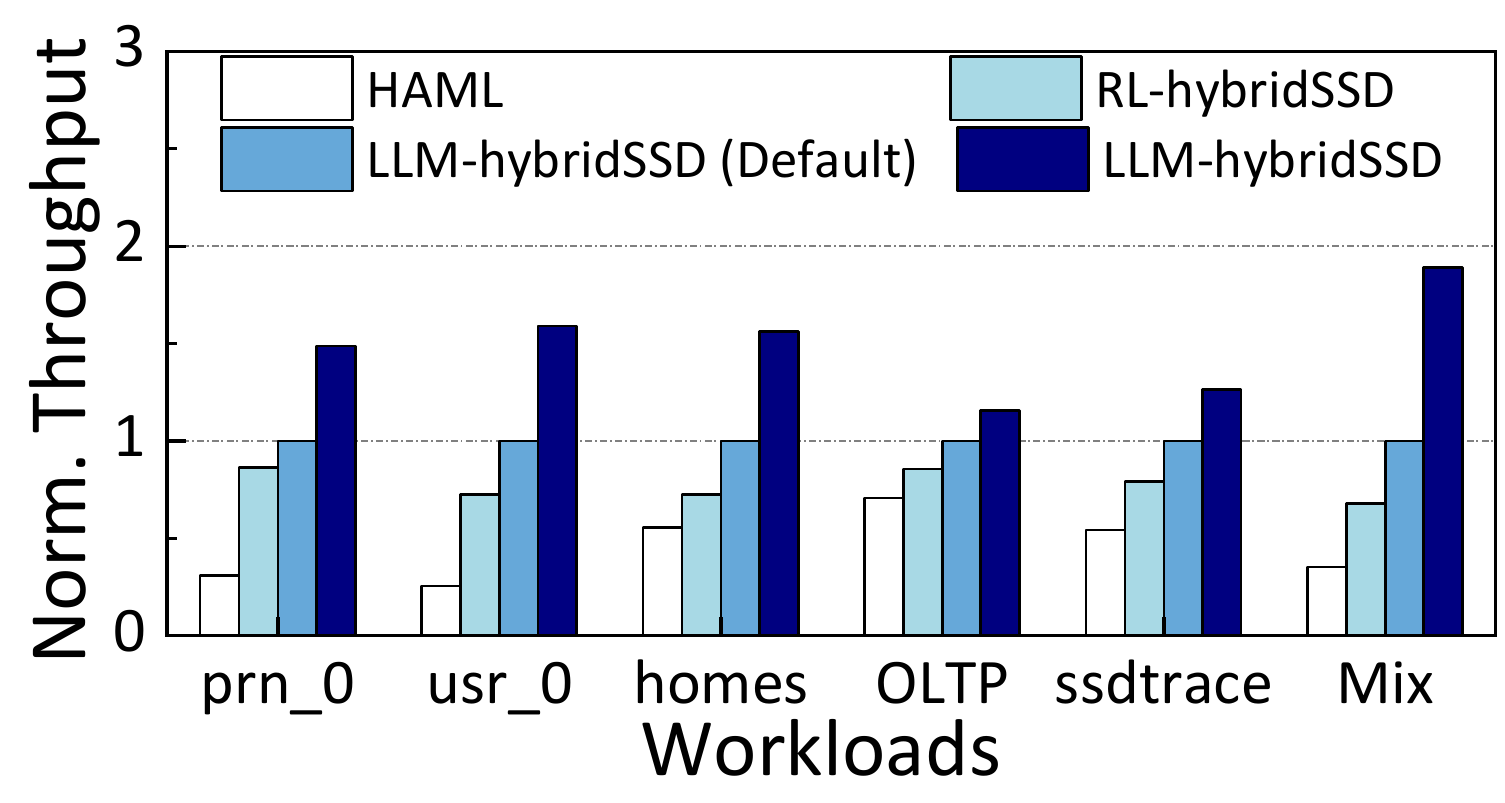} &
		\hspace{-4mm}
		\includegraphics[width=0.247\textwidth]{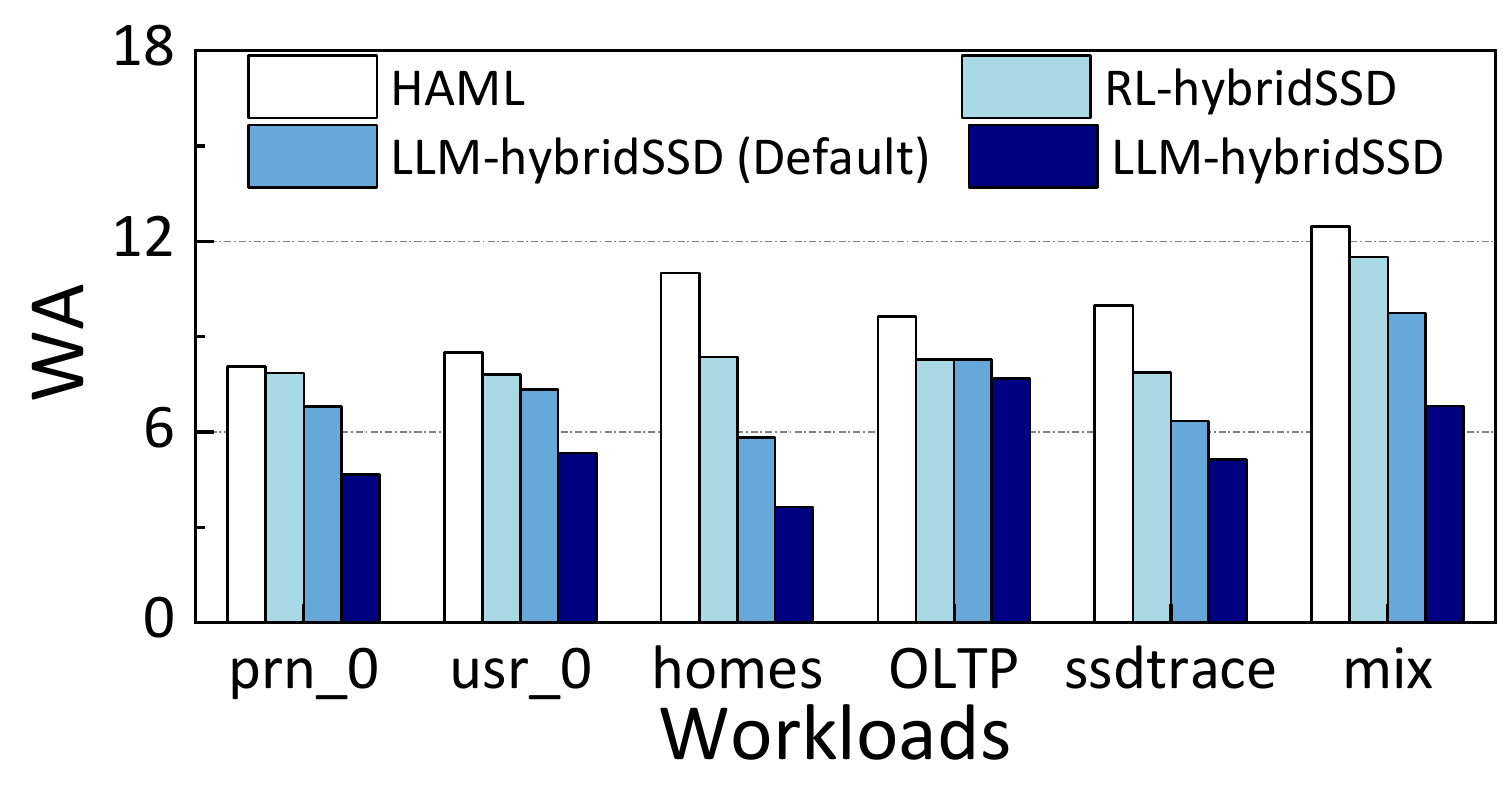} \\
  
		\hspace{-2mm}(a) Normalized throughput. &  (b) WA Coef. \\
	\end{tabular}
    %\vspace{-0.1cm}
	\caption{\label{pic:overall}Overall performance with different workloads.}
     \vspace{-0.2cm}
\end{figure}

We first evaluate the overall performance of the four implementations in terms of total throughput (normalized to LLM-hybridSSD (default)) and WA. The results are shown in Figure~\ref{pic:overall}.

$\bullet${ Throughput.} As shown, LLM-hybridSSD outperforms other strategies across the board. On average, the throughput improvements of LLM-hybridSSD over HAML, RL-hybridSSD, and LLM-hybridSSD (default) are 284.85\%, 96.17\%, and 58.92\%, respectively. Notably, for the mixed trace with rapidly changing access patterns, the throughput improvement reaches up to 88.89\%. This improvement are attributed to the adaptive parameters tuning capabilities of the LLM-based framework, which dynamically aligns configurations with hybrid SSD and trace characteristics.

$\bullet${ WA.} LLM-hybridSSD achieves significant reductions in WA, on average 43.39\%, 35.25\% and 28.56\%  compared with HAML, RL-hybridSSD and LLM-hybridSSD (default), respectively. This improvement mainly stems from the LLM-guided adjustment of data placement and migration policies, which effectively reduces unnecessary cross-mode migration and mitigates frequent GC operations.

%The results suggest that more complex models are better at managing the parameter space of hybrid SSDs.

%the API-call approach achieves greater performance gains compared to local deployment, with an average improvement of 9.66\%.  In terms of API call method, GPT-4 and Claude 3.5 achieve similar performance levels. In contrast, the locally deployed Deepseek-R1 makes highly variable parameter adjustments during optimization, leading to significant performance fluctuations and comparatively poor results compared to llama 3.2.

\subsection{Sensitivity Analysis}

\begin{figure}[t]
    \centering
    \includegraphics[width = 3.4in]{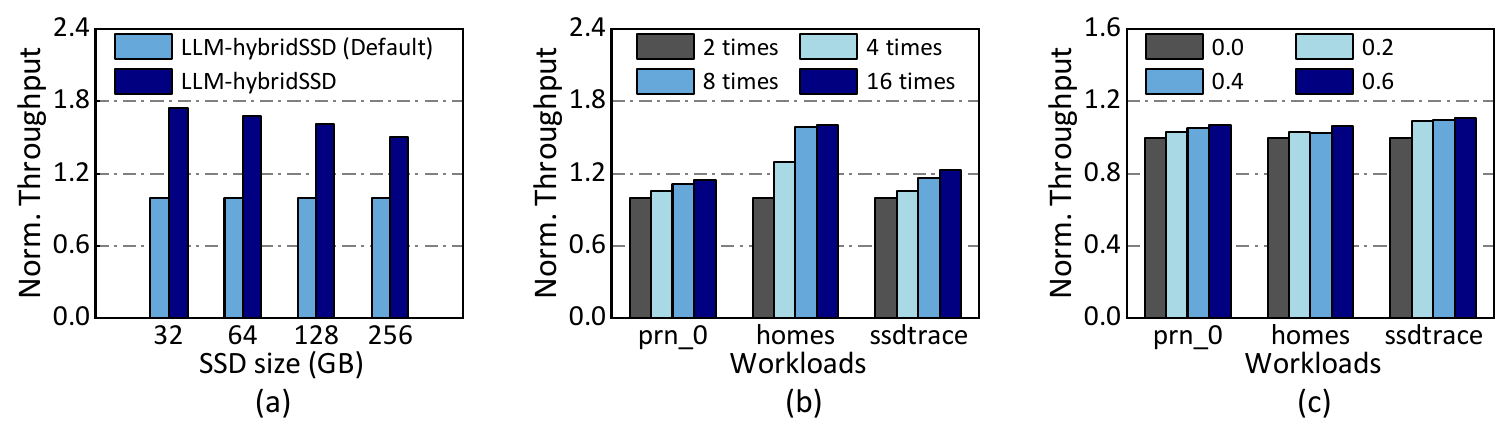}
     \vspace{-0.5cm}
    \caption{Sensitivity analysis by varying (a) SSD size, (b) tuning iterations, and (c) model temperature.}
    \label{sentivity}
     \vspace{-0.3cm}
\end{figure}

$\bullet${ Vary the SSD size:} We conduct experiments with SSD settings of different sizes using prn\_0 as a sample trace and the results are shown in Figure~\ref{sentivity} (a). As shown, the performance improvement slightly declines as SSD capacity increases. This is because larger SSDs accommodate more data, leading to fewer GC and mode conversion operations and thus less optimization potential.

$\bullet${ Vary the tuning iterations:} We conduct experiments with different tuning iteration numbers and the results are shown in Figure~\ref{sentivity} (b). As shown, increasing the number of iterations generally improves system performance, with the average throughput increasing by 32.67\%. Most gains occur in the early iterations as key parameters are rapidly aligned with workload characteristics, while improvements beyond 8 iterations gradually level off.

%Significant performance gains in the smaller number of iterations are observed as key parameters are refined to better align with workload characteristics. Beyond eight iterations, the rate of improvement begins to diminish, reflecting a plateau in performance gains.

$\bullet${ Vary the temperature of model:} We evaluate the effect of temperature settings from 0.0 to 0.6, as shown in Figure~\ref{sentivity}(c). Lower temperatures produce more deterministic outputs, stabilizing configuration selection and increasing throughput. In contrast, higher temperatures improve exploration of alternative configurations, yielding an average performance gain of 7.74\%.

\subsection{Performance Impact of different LLMs~\label{5.4}}

\begin{figure}[t]
	\centering
	\small
	\begin{tabular}{cc}
		\hspace{-5mm}	
        \includegraphics[width=0.24\textwidth]{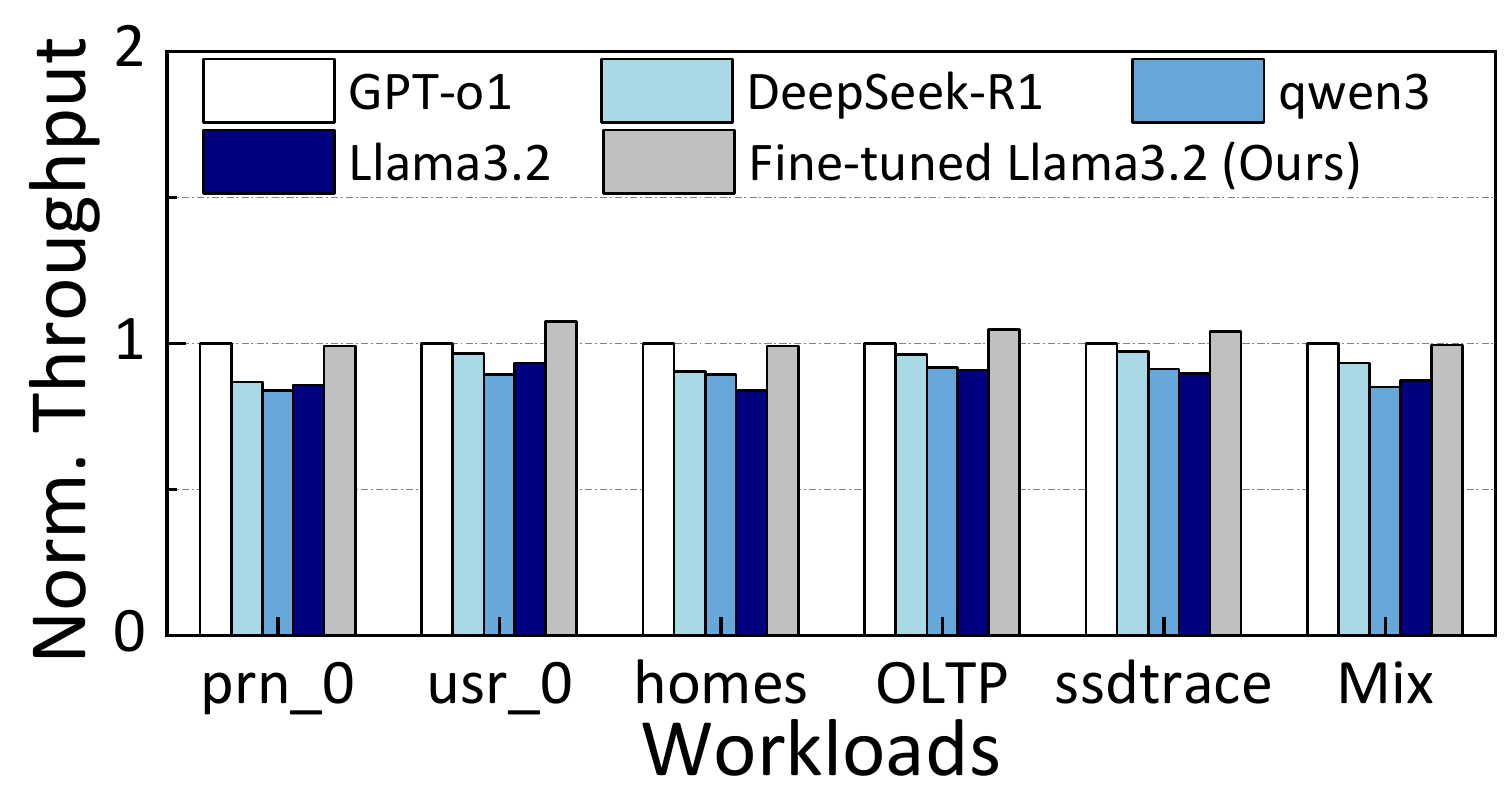} &
		\hspace{-4mm}
		\includegraphics[width=0.247\textwidth]{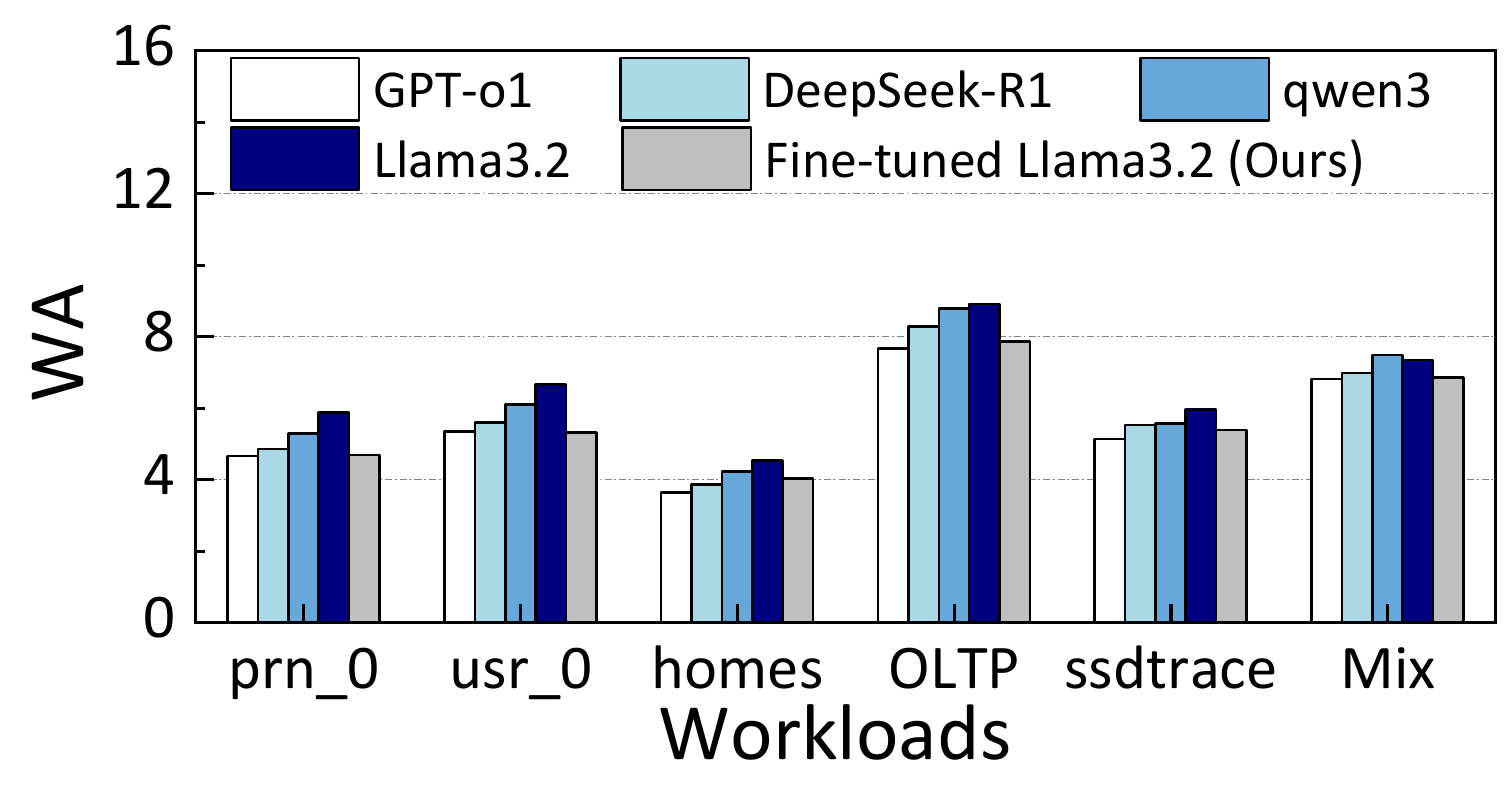} \\
  
		\hspace{-2mm}(a) Normalized throughput. &  (b) WA Coef. \\
	\end{tabular}
    \vspace{-0.2cm}
	\caption{\label{fig:model}Performance impact of using different LLM models in LLM-hybridSSD.}
     \vspace{-0.2cm}
\end{figure}

We further evaluate the impact of different LLM models on hybrid SSD optimization, including GPT-o1, DeepSeek-R1, Qwen3, LLaMA 3.2, and the fine-tuned LLaMA 3.2. Among them, GPT-o1 and DeepSeek-R1 are accessed through cloud services to explore scalability under remote inference, while Qwen3, LLaMA 3.2, and the fine-tuned LLaMA 3.2 are deployed locally to ensure lower cost and better data privacy. Figure~\ref{fig:model} presents the performance comparison across these models.

As shown, the fine-tuned LLaMA 3.2 delivers the best or comparable performance among all tested models, showing both higher throughput and reduced WA across various workloads. Compared with large-scale cloud-based models (GPT-o1 and DeepSeek-R1), the fine-tuned LLaMA 3.2 used in our architecture achieves comparable or even superior optimization performance. Meanwhile, compared with LLaMA 3.2 and Qwen3, the fine-tuned model shows better adaptation to workload characteristics and more stable configuration tuning, resulting in an average throughput improvement of 16.85\% and a WA decrease of up to 13.25\%. This improvement primarily comes from the fine-tuned model ability to internalize hybrid SSD domain knowledge, enabling it to make more accurate configuration decisions under diverse workloads.

\subsection{Overhead Analysis\label{sec:5.5}}

% used in hybrid SSD management
Table~\ref{tab:overhead} compares the inference latency and cost of different LLMs under prn\_0 workload. Cloud-based models such as GPT-o1 and DeepSeek-R1 incur the highest latency (24.37s and 69.07s, respectively) due to remote inference and network communication. In contrast, local deployed lightweight models achieve much lower latency, with Qwen3 and the fine-tuned LLaMA 3.2 recording only 20.97s and 18.92s on CPU, and 3.06s and 2.48s on GPU, respectively. Moreover, cloud-based models introduce additional monetary costs per query, whereas local deployment eliminates credit-based charges and offers greater control. The fine-tuning and sensitive-parameter identification overhead together require about 8.6 hours, paid once offline and fully amortized over long-term deployment.

%The API call method includes both network latency and cloud inference latency, whereas the local deployment method only involves CPU inference latency. The results show that even for a small model like Llama (3B parameters), local CPU inference latency remains higher than that of the API call approach. Additionally, the inference latency for Deepseek-R1 exceeds the acceptable range due to its longer chain-of-thought reasoning process.

%Table~\ref{tab:overhead} presents the overhead associated with different LLMs, including latency and cost. 

\begin{table}[]
\caption{Overhead analysis of different LLMs.}\label{tab:overhead}
 \vspace{-0.2cm}
\resizebox{0.47\textwidth}{!}{
\begin{tabular}{lcccc}
\hline
           & \multicolumn{2}{c}{\textbf{Cloud-based}} & \multicolumn{2}{c}{\textbf{Local deployed}} \\
           & \textbf{GPT-o1} & \textbf{Deepseek-R1} & \textbf{qwen3} & \textbf{Tuned LLaMA 3.2} \\ \hline
Parameters        & 200B  & 671B  & 4B    & 3B    \\
\multirow{2}{*}{LLM inference Latency (s)} 
                  & \multirow{2}{*}{24.37} 
                  & \multirow{2}{*}{69.07} 
                  & 20.97 (CPU) & 18.92 (CPU) \\
                  &        &        & 3.06 (GPU)  & 2.48 (GPU) \\
Cloud model credit Cost (\$) & 3.09  & 0.51  & /     & /     \\ \hline
\end{tabular}}
\end{table}

%CPU Usage  & /                                  & /                                       &    29.83\%            &     22.3\%                                \\
%Mem Usage  & /                                  & /                                       &     3.39GB                &     2.85GB                  \\ \hline 
%\input{7_related}
%\section{Discussion}\label{sec:Discussion}

%\textbf{Possibility of tuning in SSD controller.} We further discuss the feasibility of deploying the LLM-based tuner directly within the SSD controller rather than relying on the host. The results indicate that the runtime memory requirement for Llama 3.2 with 3B parameters is only 2.85 GB, which is feasible for deployment on an SSD controller. However, challenges remain regarding performing inference efficiently with limited computational resources and ensuring that existing SSD management policies are not disrupted. Techniques such as quantization, pruning, and lightweight runtime scheduling algorithms can effectively reduce the computational load, memory footprint, and interference with existing SSD management policies.

\section{Conclusion}\label{sec:conclusion}

In this paper, we investigate the potential of LLMs to address the challenges associated with managing hybrid SSDs. By leveraging the context-aware reasoning capabilities of LLMs, we develop the LLM-hybridSSD framework, which incorporates a fine-tuned LLM-based auto-tuner to tune the parameters of hybrid SSD management. Our experimental results demonstrate that LLM-hybridSSD achieves significant performance improvements.

%, including reductions in response time and WA when compared to traditional ML-based methods and default configuration approaches.

\section{Acknowledgments}

The work described in this paper is partially supported by the National Natural Science Foundation of China under Grant 624B2091, 62272271, U24B20149, the Shandong Provincial Natural Science Foundation under Grant ZR2024LZH004 and the Taishan Scholars Program under Grant tsqn202408009.

\bibliographystyle{ACM-Reference-Format}
\bibliography{references}

\end{document}